\documentclass{article}

\usepackage{arxiv}

\usepackage[utf8]{inputenc} % allow utf-8 input
\usepackage[T1]{fontenc}    % use 8-bit T1 fonts
\usepackage{hyperref}       % hyperlinks
\usepackage{url}            % simple URL typesetting
\usepackage{booktabs}       % professional-quality tables
\usepackage{amsfonts}       % blackboard math symbols
\usepackage{nicefrac}       % compact symbols for 1/2, etc.
\usepackage{microtype}      % microtypography
\usepackage{lipsum}		% Can be removed after putting your text content

\usepackage{graphicx}
\usepackage{epstopdf}
\usepackage{pgfplots}
\usepackage{tikz}
\usepackage{natbib}
\usepackage{doi}

\title{Testable Designs of Toffoli Fredkin Reversible Circuits}

\author{ \href{https://orcid.org/0000-0002-7952-4899}{\includegraphics[scale=0.06]{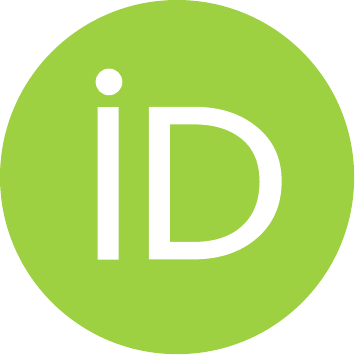}\hspace{1mm}H. M. Gaur}\thanks{Dr. Gaur is working as Associate Professor} \\
	Department of Electronics \& Communication Engineering\\
	ABES Institute of Technology Ghaziabad\\
	Uttar Pradesh, India \\
	\texttt{leoharimohan84@gmail.com} \\
	\And
	{A. K. Singh} \\
	Department of Computer Applications\\
	National Institute of Technology Kuruskshetra\\
	Haryana, India \\
	\texttt{ashutosh@nitkkr.ac.in} \\
	\And
{U. Ghanekar} \\
Electronics \& Communication Engineering\\
National Institute of Technology Kuruskshetra\\
Haryana, India \\
\texttt{ugnitk@nitkkr.ac.in} \\
}

\renewcommand{\shorttitle}{\textit{arXiv} Template}

\hypersetup{
pdftitle={A template for the arxiv style},
pdfsubject={q-bio.NC, q-bio.QM},
pdfauthor={David S.~Hippocampus, Elias D.~Striatum},
pdfkeywords={First keyword, Second keyword, More},
}

\begin{document}
\maketitle

\begin{abstract}
Loss of every bit in traditional logic circuits involves dissipation of power in the form of heat that evolve to the environment. Reversible logic is one of the alternatives that have capabilities to mitigate this dissipation by preventing the loss of bits. It also have the potential to broaden the horizon of futuristic reckon with its applications to quantum computation. Application of testing strategies to the logic circuits is a necessity that guarantees their true functioning where the researchers are at par with solutions for the upcoming challenges and agreements for reversible logic circuits. Novel methods of designing Toffoli, Fredkin and mixed Toffoli-Fredkin gates based reversible circuits for testability are put fourth in this article. The proposed designs are independent of the implementation techniques and can be brought into real hardware devices after obtaining a stable fabrication environment. The experimentation for the proposed models are performed on RCViewer and RevKit tools to verify the functionality and computation of cost metrics. Fault simulations are carried out using C++ and Java to calculate fault coverage in respective methodologies. The results confirmed that all the presented work outperforms existing state-of-art approaches.
\end{abstract}

% keywords can be removed

\section{Introduction}
\begin{figure}[!h]
	\centering
	\includegraphics[width=85mm]{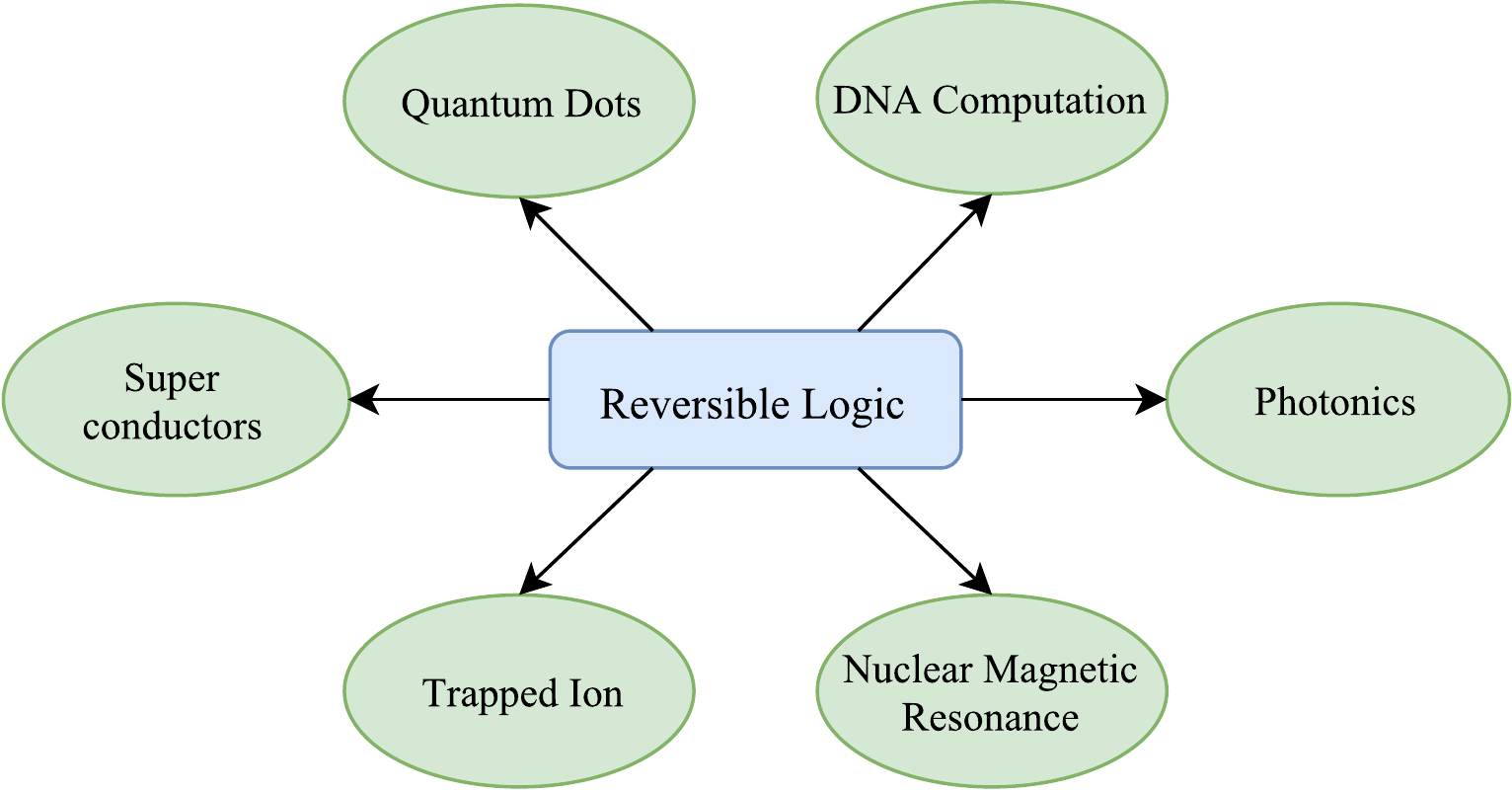}
	\caption{Computing technologies} 
	\label{figintro}
\end{figure}

In the current scenario, electronic industries are facing the problems of power utilization and overheating of equipment. In the past decades, these issues were used to be solved by adopting the practice of reducing the size of traditional transistors which has also been miniaturized  up to certain nanometres. However, if the size of the transistors are further scaled, the power and overheating problems will increase exponentially \cite{ieeespectrum,ieeespectrum1}. Moreover, the well-known Landauer's theory extricates the limitations of irreversible computation which also bounds the size of transistors in conventional logic circuits as they also involve loss of information in the form of heat \cite{landuer}. Meanwhile, the demands of more and more applications in single SOCs are also leading to a drastic increase in loss of information. Reversible logic is one of the promising techniques to reduce the power requirements, as these circuits are theoretically proven for providing nearly energy free computation by preventing the loss of information and have the capability of producing ultra high speed and compact electronic devices \cite{Bennett:1973:LRC:1664562.1664568}. However, the logic can be applied to traditional logic circuits, but its applications to quantum computation have been proven for achieving excellence in terms of power consumption, speed and size \cite{quantumbookNielsen}.  The identification and implementation of reversible quantum circuits have also been achieved using several probabilistic methods and ideas. Fig. \ref{figintro} shows some of the dominating technologies where the researchers are currently exploring the possibilities for employing this logic at physical foregrounds \cite{selfintro}.

The framework of reversible logic circuits design and synthesis techniques is based on Toffoli and Fredkin gates, which can be further scaled into $n$-$th$ order gates and libraries, commonly known as Multiple Control Toffoli (MCT) and Multiple Controlled Fredkin (MCF). Several other gates have also been proposed in the literature, but the primary components of these gates are MCF and MCT. Moreover, the final quantum decomposition of the reversible circuits are based on them \cite{selftandf}. The efficiency of the designs are governed by several performance metrics defining their operating cost. These metrics are number of wires, gate cost, quantum cost and garbage output. Testing has also been extensively studied since last decade for the recognition of several types of fault models in reversible circuits. A number of novel paradigms have been presented in both the area of online and offline testing of reversible logic circuits. Online testable environment are provided over pristine design methodologies and circuit modification principles. Test data minimization in offline tesing is achieved over new deterministic, randomized test patter generation algorithms and circuit modification techniques for respective faults.  The reduction of operating cost has been achieved to some needful extent in all the proposed approaches with respect to prior ones for narrowing the compensation with overall testing overheads \cite{selffaultmodels}.

\begin{figure*}[!h]
	\centering{	\includegraphics[width=150mm]{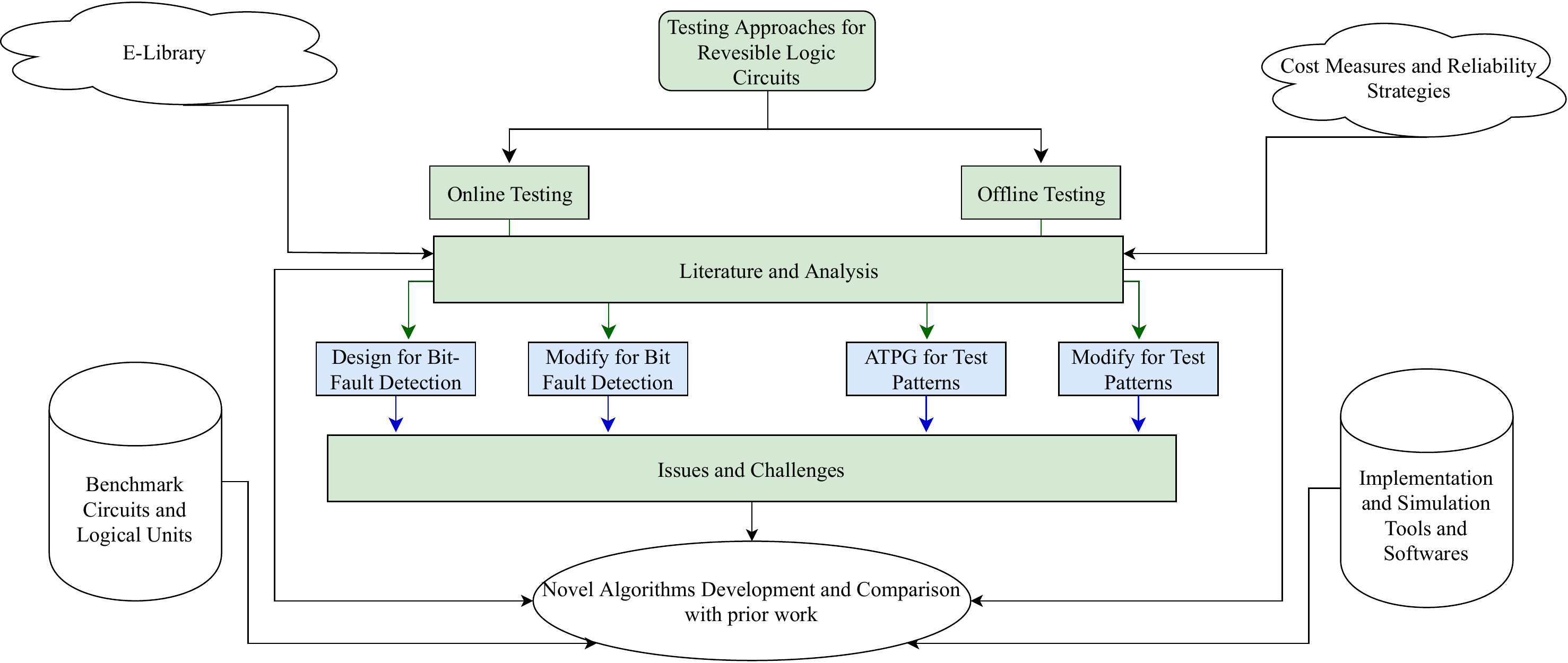}}
	\caption{Framework for designing testable reversible circuits} 
	\label{figgen}
\end{figure*}

A comprehensive and comparative analysis of the existing online and offline testing methodologies for nearly all fault models in reversible circuits has been completed in correlation with the problem statement \cite{selfprocediareviewonline,selfpertanika,selfintegration,selffaultmodels} in the beginning of the proposed work. An overview of reversible logic, cost metrics and associated faults models are also explained for providing background of the work. Overall work in the literature and deeply analyzed and organized into four set of categories that defining the plan of action for the novel development in the area. The illustration in Fig. \ref{figgen} which shows a generalized framework to achieve the quoted objectives. As overall  framework is based on fundamental MCT and MCF gates. At first, an in-depth and comparative analysis of nearly all reversible gates  has been done. A three level analysis i.e., gate, design and testability level, has been performed to confirm the efficacy of the fundamental gates \cite{selfprocediagates,selftandf}. At last, the proposed testable design methodologies for online testing are also applied to obtain an efficient set of testable Data Path Elements (DPE) designs.

The objectives of the work described in this paper from the authors thesis \cite{phdthesis} is to development of testable design methodologies at reduced testing overheads in terms of reversible circuit cost metrics, test data volume, design complexity and time. There are the following contributions toward the achievement of stated objectives:

\begin{itemize}
	
	\item Novel design methodologies using Multiple Controlled Toffoli (MCT), Multiple Controlled Fredkin (MCF) and mixed Multiple Controlled Toffoli-Fredkin (MCTF) gates which shows built-in testability features towards single bit faults. 
	
	\item New circuit modification methodologies for MCT, MCF and MCTF circuits are introduced for the detection of single bit faults. 
	
	\item Efficient circuit modification methodologies along with general test sets are proposed for MCT, MCF and MCTF circuits for the detection of stuck-at faults to minimize the volume of test data.
	
	\item New testable designs of Full Adder, Ripple Carry Adder, $4$-bit reversible array based Multiplier and Arithmetic \& Logic Unit are proposed using MCT and MCF gates.
\end{itemize}

Lining up with the targeted objectives, a detailed overview and analysis of the proposed work is described in the consecutive sections:

\section{Circuit Design and Modification Methodologies for Online Testing}
\begin{figure*}[!h]
	\centering
	\includegraphics[width=150mm]{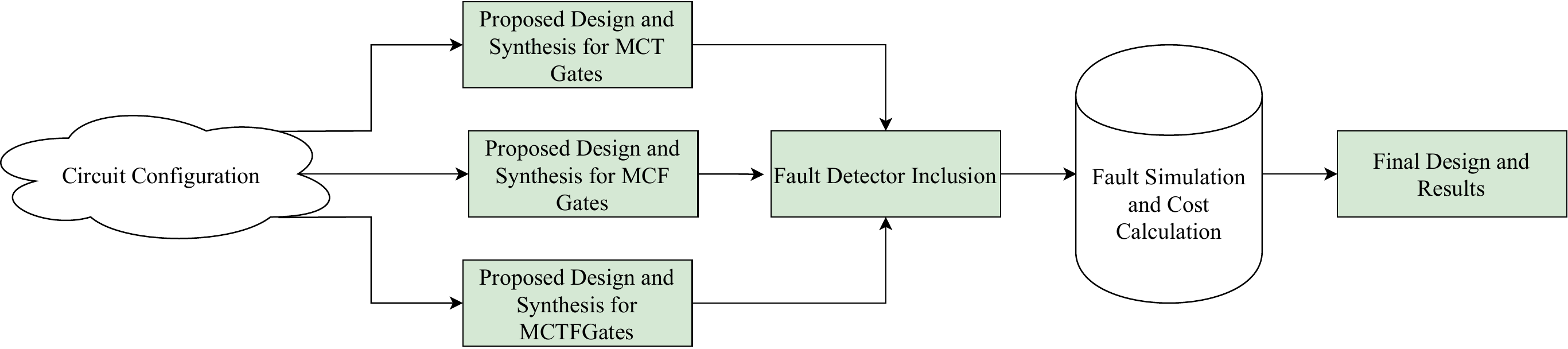}
	\caption{Synthesis and test flow of design methodologies for online testing} 
	\label{figm1}
\end{figure*}

Extensive design methodologies are realized for the construction of MCT, MCF and MCTF circuits. The constructed circuits provide built-in testability feature for the detection of single bit faults. As the utilization of parity preserved architecture with arbitrary design methodology ensures the detection of single bit flip faults in logic circuits. The methodologies are engaged in the creation of parity preserving circuits using MCT, MCF and mixed MCTF gates followed by the fault detection process.  The design and test flow of the formulation of these methodologies is depicted in Fig. \ref{figm1}.
First, an MCT gates placement technique is proposed for producing parity preserving circuits \cite{selfiet}. Second, the properties of MCF gates are exploited to showcase the scheme for the detection of faults \cite{selfmcfonlineoffline}. Third, the MCT gate placement method is utilized in combination with MCF gates to achieve testability in MCTF circuits  \cite{selfijpap}. The fault detection is achieved by cascading a parity checker in the circuit using CNOT gates from the inputs and outputs to an additional wire. The circuits produced using proposed methods are incorporated with testability feature rather put extra efforts in converting original circuit into their testable form. A set of benchmark circuits and corresponding testable designs are implemented to observe the cost of designing and proving the efficacy of proposed schemes over existing ones. 

\begin{figure*}[!h]
	\centering
	\includegraphics[width=150mm]{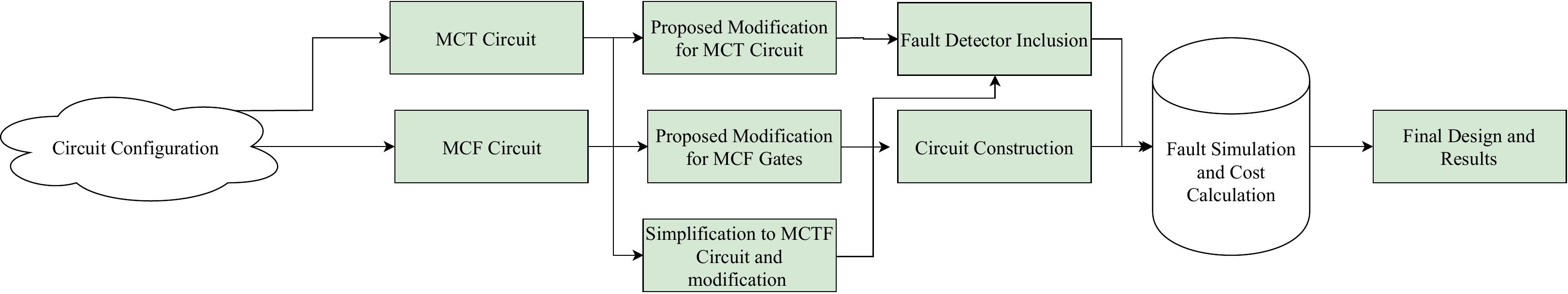}
	\caption{Design and test flow of modification methodologies for online testing} 
	\label{figm2}
\end{figure*}

%\subsection{Circuit Modification Methodologies for Online Testing}
Modifications for testability in logic circuits accounts a large increment in operating cost which enhances overall cost of manufacturing. New modification schemes for MCT, MCF and mixed MCTF circuits are   introduced at lower operating cost. The modification procedures utilize the technique of parity preservation and generation for providing full coverage of single bit faults. The modification and test flow of the formulation and obtaining the measures for these methodologies is depicted in Fig. \ref{figm2}.

Gates cascading terminology is used for the modification of MCT circuits at the start as these gates are widely used for designing reversible circuits.  The method requires only a single wire, around twice number of gates and zero garbage cost for its formulation \cite{selfoptik}. Derived gates technique is used for the modification of MCF circuits. The gates are transformed into corresponding testable that provide fault detection as well as location functionality in MCF circuits. It utilized parity preserving gates rather to than to convert a modify a gate into corresponding parity preserving form \cite{selfieee}.  A three stage process is explored for converting MCT into MCTF circuits which largely decreases the gate cost using a wire by utilizing gates cascading and parity preserving the property of MCF gates \cite{selfmctfonline}.  The method involves the steps of  simplification of MCT circuits into MCTF cascades and modification of the resultant circuit. A large set of circuits and benchmarks are designed using proposed methodologies and fault simulations are performed to evaluate the performance and validate their functionality. The detection of single bit faults are targeted  in all the proposed methodologies.  Results prove that the present methods provides excellent reduction in the operating costs as compared to existing work in this area and provide full coverage of single bit faults. 

\section{Circuit Modification Methodologies for Offline Testing}
\begin{figure*}[!h]
	\centering
	\includegraphics[width=150mm]{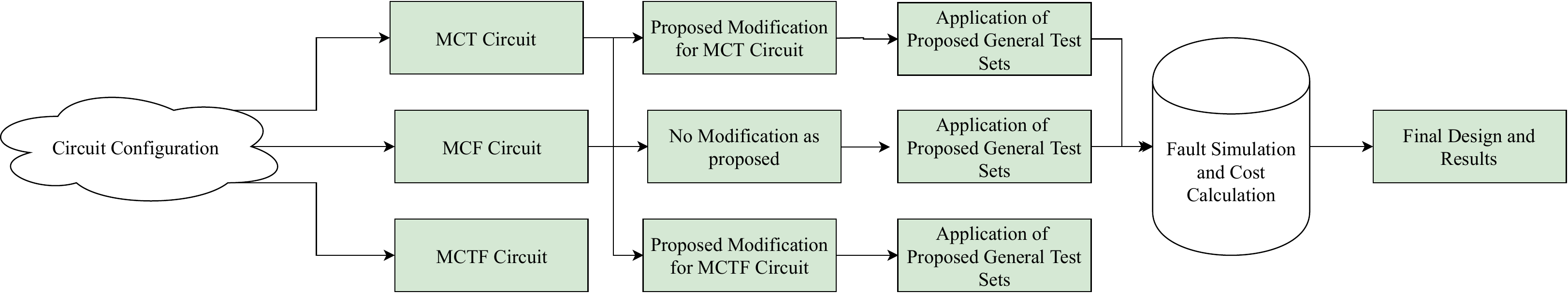}
	\caption{Design and test set application flow of modification methodologies for offline testing} 
	\label{figm3}
\end{figure*}

Test set generation in reversible circuits is followed by a number of methodologies for the detection of faults. These methodologies utilize specific deterministic ATPG, randomized ATPG and modification approaches for the detection of stuck-at, bridging, missing gate, cross-point and cell faults. The existence of the trade-off between testability and overheads can be seen in all the prior methodologies in terms of performance measures like gate cost and quantum cost, test size and time utilization. New modification methodologies  are introduced for the detection of stuck-at faults in MCT, MCF and MCTF circuits using test sets of minimal sizes. The flow for the formulation of these methods and simulations of  GTS to obtain the effective measures is depicted in Fig. \ref{figm3}. 

The MCT circuits are modified in such a manner that the applied test vector reaches all the levels without any change in values on the wires of the circuit \cite{selfdrdo}. An ($n+1$) dimensional general test set ($GTS$) containing only two test vectors is presented, which provides full coverage of single and multiple stuck-at faults in the circuit. Here $n$ denotes the number wires contained by the circuit. Deterministic approaches for the identification and detection of different types of fault models in MCF circuits are introduced \cite{selfmcfonlineoffline}. The conservative property of MCF gates is utilized for multiple types of fault detection in these circuits by the three test sets of sizes $2$, $n$ and $2(n-2)$. Moreover, both the schemes are combined for the detection of stuck-at faults in MCTF circuits \cite{selfmctfoffline}. All the methodologies have experimented on several benchmark circuits, where an excellent reduction in overall operating costs has been achieved as compared to prior work experimented in the same platform. Moreover, these circuits can be tested by general test sets of fixed sizes without spending the excess time required to formulate specific algorithm under stuck-at fault detection. 

\section{Testable Designs of Data Path Elements}
\begin{figure*}[!h]
	\centering
	\includegraphics[width=140mm]{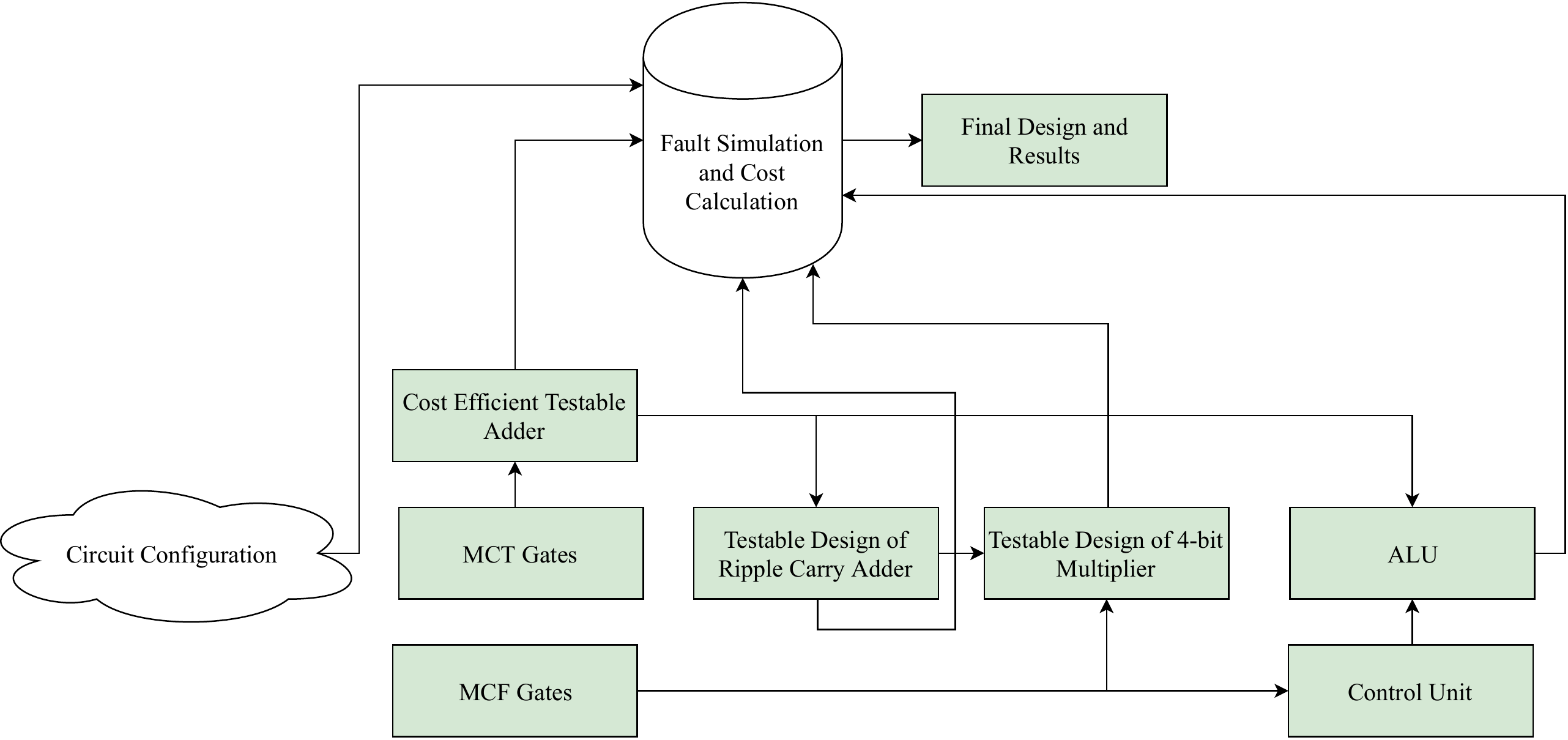}
	\caption{Design and Implementation flow of DPE} 
	\label{figdpe}
\end{figure*}

Modern digital processors are comprised of several data path elements (DPE) like adders, multipliers, multiplexers, logical shifters, arithmetic logic unit etc. These elements are the functional units within the microprocessor which are used to execute computational operations. New testable architectures of a full adder (FA), ripple carry adder (RCA), multiplier (MUL) and an arithmetic \& logic unit (ALU)  using MCT and MCF gates \cite{selfmulandadde,selfalu}. The design and simulation flow of these elements is shown in Fig. \ref{figdpe}. 

The major role is played by the creation of parity preserving circuits for incorporating testability in overall circuits realization. Firstly a full adder is created which is used to develop the architecture of RCA. A $4$-bit reversible array based multiplier with scalability factor of order $4N$ by using RCA and Fredkin gates \cite{selfmulandadde}. Design of Arithmetic Logic Unit (ALU) which can be scalable up to $N$ number of bits is proposed \cite{selfalu} by combining a novel structure of control unit (CU) using Fredkin gates and FA. that . The ALU design can be scalable for $N$-bit operations. These designs can be scalable for $N$-bit operations  and incorporates testability features for the detection of single-bit faults at lower overheads which shows their exclusive features. The superiority of the designed circuits is acknowledged by implementing them using reversible circuit analyzer tool and obtaining corresponding operating costs. 

\section{Results and Assessment}
The experiments for evaluating the efficacy of the proposed work in this thesis were performed on a machine with 64-bit Ubuntu-16.04LTS having Intel Core I7-4790, 3.60 GHz clock and 4GB memory. The prerequisites which can be seen in different part of the thesis are listed as follows:
\begin{itemize}
	\item The benchmark circuits description in the form of \textit{pla} and \textit{tfc} are taken from reversible logic synthesis and benchmark pages \cite{benchmaslov,revlib}.
	\item Revkit-A toolkit is used for reversible logic synthesis \cite{revkit}.
	\item The circuits are synthesized using well-known garbage free transformation based synthesis algorithm.
	\item RC-viewer and RC-Viewer+ tools for designing reversible circuits is used for calculating respective measures that define operating costs \cite{benchmaslov}.
	\item QCA-designer is used for the implementation proposed designs and finding out physical reliability at QCA level. \cite{walus2004qcadesigner}.
	\item Fault coverage is verified using simulations carried out in C++ and Java.
\end{itemize}

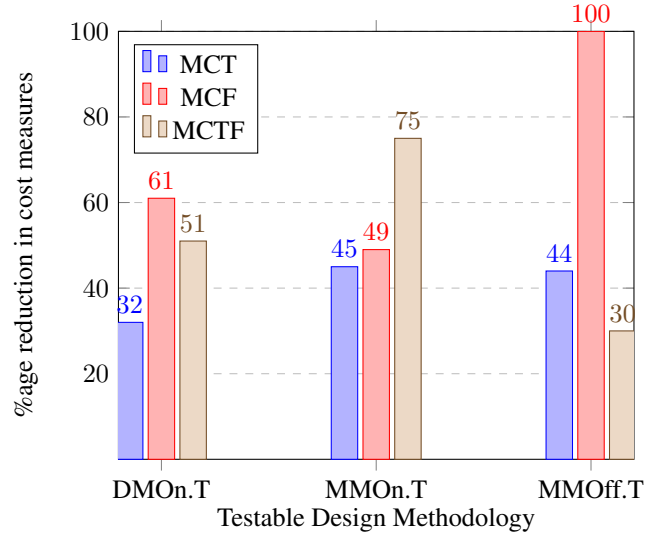
\begin{figure}[!h]
	\centering
	\begin{tikzpicture}
	\begin{axis}[
	xlabel={Testable Design Methodology},
	ylabel={\%age reduction in cost measures},
	ybar,
	ymin=0, ymax=100,
	ytick={20,40,60,80,100},
	symbolic x coords={DMOn.T,MMOn.T,MMOff.T},
	xtick=data,
	nodes near coords,
	legend pos=north west,
	ymajorgrids=true,
	grid style=dashed
	]
	\addplot table[x=interval,y= MCT]{\mydata};
	\addplot table[x=interval,y=MCF]{\mydata};
	\addplot table[x=interval,y=MCTF]{\mydata};
	\legend{MCT,MCF,MCTF }
	\end{axis}
	\end{tikzpicture}
	\caption{Result analysis of presented testing methodologies}
	\label{con123} 
\end{figure}

A large set of benchmark circuits are taken from the two platforms which are experimented on the basis of each design methodologies. Design, synthesis and implementation of the methodologies are done using RC Viewer and Revkit tools. QCA structures are also implemented in  to obtain the cost measures in some of the cases. Fault simulation is done by creating programs that realize the circuits and computing the fault coverage after inducing a type of fault for which the design method has been developed. The implementation and simulation of the prior methodologies in the domain has also been executed to compare the presented work and calculate the efficacy. 
It has been analyzed the present work in the domain of design methodologies for online testing (DMOnT) has achieved the maximum reduction in cost measures by $12\%$  in case of MCT based designs, $61\%$ in case of MCF based approaches and $51\%$ in case of MCTF based design methodologies. The proposed work in the domain of modification methodologies for online testing (MMOnT) has achieved the reduction in cost measures by $45\%$ in MCT based modification techniques, $49\%$ in case of MCF based techniques and $75\%$ in case of MCTF based modification methodologies.  The work done in the domain of modification methodologies for offline testing (MMOffT) has achieved the reduction in cost measures up to $44\%$ in MCT based techniques, $100\%$ in case of MCF based techniques and $30\%$ in case of MCTF based methodologies. These analytics also pictured in Fig. \ref{con123}. 

\pgfplotstableread[row sep=\\,col sep=&]{
	interval & MCT       & MCF & MCTF   \\
	DMOn.T            & 32      & 61  & 51    \\
	MMOn.T      & 45      & 49  & 75    \\
	MMOff.T              & 44      & 100  & 30  \\
}\mydata

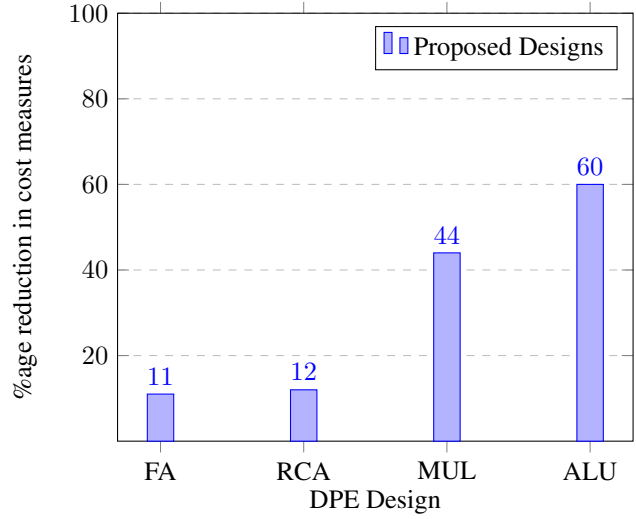
\begin{figure}[!h]
	\centering
	\begin{tikzpicture}
	\begin{axis}[
	xlabel={DPE Design},
	ylabel={\%age reduction in cost measures},
	ybar,
	ymin=0, ymax=100,
	ytick={20,40,60,80,100},
	symbolic x coords={FA,RCA,MUL,ALU},
	xtick=data,
	nodes near coords,
	legend pos=north east,
	ymajorgrids=true,
	grid style=dashed
	]
	\addplot table[x=interval,y=Proposed]{\mydata};
	\legend{Proposed Designs }
	\end{axis}
	\end{tikzpicture}
	\caption{Result analysis of presented DPE architectures}
	\label{con444} 
\end{figure}

The performance of the presented DPE structures is analyzed by implementing $4$-$64$ bit circuits over reversible circuit analyzer tool and compare the characteristics with the prior efficient architectures . Reported results showing the reduction of cost measures are illustrated in Fig. \ref{con444}. Presented testable FA circuits achieved an average reduction by $11\%$ in the when all the considered parameters are combined together. RCA achieved a reduction by $12\%$ and $44\%$ has been achieved in case of MUL. A reduction up to $60\%$ in the gate cost has been achieved with respect to recently reported reversible ALU architectures from the literature. 

\pgfplotstableread[row sep=\\,col sep=&]{
	interval & Proposed   \\
	FA       & 11    \\
	RCA      & 12    \\
	MUL      & 44  \\
	ALU      & 60  \\
}\mydata

%%%%%%%%%%%%%%%%%%

The fault coverage is also calculated in nearly all the proposed methodologies and designs. Full coverage has been achieved in the respective methodologies for a considered fault model. The calculations show a large reduction in operating costs when compared to the prior work in all designs and testing methodologies. The requirements of extra hardware and time to attain testability can be eliminated by utilizing these methodologies during the design process. Hence, the methods provide solutions to both the problem of designing and testability of reversible circuits which can be adopted by any synthesis algorithm to minimize testing overheads.

\section{Conclusion and Future Scope}
The change in technology will give rise to new challenges where the manufacturers would have to commence certain proofs for their truthful functionality to be deliverable in the huge market containing highly ambitious consumers. Testing is the only way out for them to rescue from these situations. However, it is a necessary exercise but it deals with a large increment in operating costs. Moreover, a huge amount of power consumption is governed by the testing methodologies used by the manufacturers. Numerous approaches for constructing built-in testable MCT, MCF and mixed MCTF circuits over novel design methodologies and circuit modification techniques are presented for the detection of single bit faults. The performance of all the approaches is analyzed by experimenting on a set of benchmark circuits. As the logic circuits are very much prone to the occurrence of stuck-at faults, new circuit modification techniques for minimization of test data in MCT, MCF and mixed MCTF reversible circuits are introduced for their detection. In addition, this work introduces new testable designs of scalable adders, multiplier and arithmetic logic unit for future microprocessors. MCT and MCF gates are taken into account for the formulation of all the proposed approaches as they are proven universal as well as superior for designing and testing of reversible circuits. The efficacy of all the modules is justified by providing the implementation on reliable tools for reversible circuits. The fault tolerance designing model has been identified that utilized the proposed methods in this paper \cite{selffaulttolerance}, however, some of the possible limitations and scope still exists: 

\begin{itemize}
	\item This work is merely a start to the research needed in determining the feasibility of reversible circuits as a replacement to present CMOS technology.
	
	\item Development of a comprehensive tool for synthesizing reversible circuits based on the proposed framework.
	
	\item Efficient ATPG Algorithms can be explored to minimize the test-data volume for the detection of multiple types of fault models in reversible circuits.
	
	\item MCF gates based synthesis algorithm can be developed, as it has not gained significant attention by the researchers working in this area.
	
\end{itemize}

\section*{Acknowledgment}
Sincerest thanks to all the reviewers for their  extensive and insightful comments and suggestions on the thesis and supported manuscripts. Each and every comment on the manuscript motivated the authors to perform better. Special thanks to Dr. Jimson Mathew, Associate Professor and Head, Department of Computer Science \& Engineering, IIT Patna, for his excellencies during the final defense of work.

%Photons, trapped atoms, n
%\begin{gather}
%f(k_m,T)=(k_{PR})\oplus T
%\label{eq.MCT}
%\end{gather} 
. 

\bibliographystyle{unsrtnat}
\bibliography{myref}  
\end{document}